\documentclass[%
preprint,
showpacs,preprintnumbers,
 amsmath,amssymb,
 aps,
 prc,
 longbibliography,
]{revtex4-1}

\usepackage{graphicx}
\usepackage{dcolumn}
\usepackage{bm}
\usepackage{hyperref}

\begin{document}


\title{New integral formula and its applications to light nucleus reactions}


\author{Xiaojun SUN}
\email{sxj0212@gxnu.edu.cn}
\affiliation{College of Physics, Guangxi Normal University, Guilin 541004, P. R. China;\\
State Key Laboratory of Theoretical Physics, Institute of Theoretical Physics, Chinese Academy of Sciences, Beijing 100190, P. R. China;}

\author{Jingshang ZHANG}
\affiliation{China Institute of Atomic Energy, P. O. Box 275(41), Beijing 102413, P. R. China}



\date{\today}

\begin{abstract}
A new integral formula, which has not been compiled in any integral tables or mathematical softwares, is proposed to obtain the analytical energy-angular spectra of the particles that are sequentially emitted from the discrete energy levels of the residual nuclei in the statistical theory of light nucleus reaction (STLN). In the cases of the neutron induced light nucleus reactions, the demonstration of the kinetic energy conservation in the sequential emission processes becomes straightforward thanks to this new integral formula and it is also helpful to largely reduce the volume of file-6 in nuclear reaction databases. Furthermore, taking p+$^9$Be reaction at 18 MeV as an example, this integral formula is extended to calculate the energy-angular spectra of the sequentially emitted neutrons for proton induced light nucleus reactions in the frame of STLN.

\end{abstract}


\maketitle

\section{\label{sect1}Introduction}

Nuclear reaction database contains recommended cross sections, spectra, angular distributions, fission product yields, photo-atomic and thermal scattering law data. These data were analyzed by experienced nuclear physicists to produce recommended libraries for one of the national nuclear data projects. All data are stored in the internationally-adopted ENDF-6 format maintained by the Cross Section Evaluation Working Group (CSEWG). File-6, one of the important files of nuclear reaction database, is recommended when the energy and angular distributions of the emitted particles must be coupled, when
it is important to give a concurrent description of neutron scattering and particle emission,
when so many reaction channels are open that it is difficult to provide separate reactions,
or when accurate distributions of the charged particle or residual nucleus are required for particle
transport, heat deposition, or radiation damage calculations \cite{Herman2009}.

However, the evaluated neutron nuclear data of file-6 for 1p-shell light nuclei, such as $^{6,7}$Li, $^9$Be, $^{10,11}$B, $^{12}$C, $^{14}$N and $^{16}$O, are not satisfactory in major libraries which were released recently. These libraries include FENDL-3.0 \cite{FENDL-3.02015}, TENDL-2014 \cite{TENDL-2014}, JEFF-3.2 \cite{JEFF-3.22014}, JENDL-4.0u2 \cite{JENDL-4.0u22012}, ENDF/B-VII.1 \cite{ENDF/B-VII.12011}, ROSFOND-2010 \cite{ROSFOND-2010}, and so on. The spacings of the discrete energy levels for 1p-shell light nuclei are so large that all of the sequentially emitted particles can always reach the discrete levels of their residual nuclei with large level widths. Furthermore, some residual nuclei such as $^5$He, $^5$Li, $^6$He and $^8$Be, are unstable, thus the two-body, three-body and double two-body breakup processes arise as the simultaneous particle emissions. In addition, because of light mass, the recoil effect of the energy conservation must be strictly taken into account. Therefore, the proper description of the double-differential cross sections (or the energy-angular spectra) in light nucleus reactions is a very complicated problem. There is lack of the appropriate theoretical methods for reasonably describing these distinguishing features in nuclear reaction databases mentioned above for both neutron and proton induced light nucleus reactions.

Recently, based on the previous studies \cite{Zhang2009,Zhang2011}, a statistical theory of light nucleus reactions (STLN) is proposed to set up file-6 for both neutron and proton induced 1p-shell light nucleus reactions \cite{Zhang2015}. In the frame of STLN, the contributions from the particles that are sequentially emitted between the discrete energy levels of the residual nuclei are very important to the total energy-angular spectra. For straightforwardly and analytically describing the energy-angular spectra of the secondary sequentially emitted particles, a new integral formula is proposed in STLN. This new integral formula has not been compiled in any integral tables \cite{Gradshteyn2007,Arad2002,Bradley1995} or mathematical softwares such as Wolfram Mathematica \cite{Mathematica}, Matlab \cite{Matlab}, Maple \cite{Maple}, et al.

This paper is organized as follows: In Sec. II, a new integral formula is proposed and proved. In Sec. III, the applications of the integral formula are introduced to describe the energy-angular spectra of the secondary sequentially emitted particles for both neutron and proton induced light nucleus reactions where energy conservation is strictly kept. Finally, a summary is given in the last section.

\section{Integral formula}
\label{sect2}

Assuming $\theta$ is the difference between azimuths $(\theta_1, \varphi_1)$ and $(\theta_2, \varphi_2)$, we have
\begin{equation}\label{eq1}
\cos \theta=\cos \theta_1 \cos \theta_2+\sin \theta_1 \sin \theta_2 \cos(\varphi_1-\varphi_2).
\end{equation}

In addition, the (associated) Legendre functions satisfy the following relation \cite{Gradshteyn2007}
\begin{equation}\label{eq2}
P_l(\cos \theta)=P_l(\cos \theta_1)P_l(\cos \theta_2)+2\sum_{m=1}^{l}\frac{(l-m)!}{(l+m)!}P_l^m(\cos\theta_1)P_l^m(\cos \theta_2)\cos m(\varphi_1-\varphi_2),
\end{equation}
where $P_l(x)$ and $P_l^m(x)$ are the Legendre function and the associated Legendre function, respectively.

We set $t=\varphi_1-\varphi_2$ and integrate $t$ from 0 to $\pi$, the integral formula can be easily derived as follows
\begin{eqnarray}\label{eq3}
\int_0^{\pi}P_l(\cos \theta)dt&=& \int_0^{\pi}dtP_l(\cos \theta_1 \cos \theta_2+\sin \theta_1 \sin \theta_2 \cos t) \nonumber\\
&=& \pi P_l(\cos\theta_1)P_l(\cos\theta_2).
\end{eqnarray}

Setting $\eta=\cos \theta_1$, then Eq. (\ref{eq3}) takes the following form
\begin{eqnarray}\label{eq4}
\int_0^{\pi}dt P_l(\sqrt{(1-\eta ^2)\sin^2\theta_2}\cos t+\eta\cos\theta_2)=\pi P_l(\eta)P_l(\cos\theta_2).
\end{eqnarray}

Eq. (\ref{eq4}), which has not been compiled in any integral tables or mathematical softwares, has been widely used in STLN. Using this integral formula, we can obtain the analytical energy-angular spectra of the sequentially emitted particles in the frame of STLN. In addition, it can also largely reduce the volume of file-6 in nuclear reaction databases.

\section{Applications to light nucleus reactions }
\label{sect3}
\subsection{Double-differential cross section of the secondary emitted particles and the energy conservation}

For light nucleus reactions, it is assumed that the first residual nucleus $M_1$ is at the discrete energy level $E_{k_1}$, after the first particle $m_1$ is emitted from the compound nucleus $M_C$ in the center of mass system (CMS, denoting superscript $c$). Considering the energy-momentum conservation in CMS, the definitive kinetic energies of $m_1$ and $M_1$ can be easily derived as
\begin{eqnarray}\label{eq5}
\varepsilon_{m_1}^c=\frac{M_1}{M_C}(E^*-B_1-E_{k_1})
\end{eqnarray}
and
\begin{eqnarray}\label{eq6}
E_{M_1}^c=\frac{m_1}{M_C}(E^*-B_1-E_{k_1}).
\end{eqnarray}
Where $E^*$ is the excited energy of $M_C$, and $B_1$ is the binding energy of $m_1$ in $M_C$. For convenience, all the masses also indicate the corresponding nuclei or particles in text. It is obvious that there is an approximate relation $M_C\approx m_1+M_1$ without lowering the calculated precision.

The normalized angular distributions of $m_1$ and $M_1$ with definitive kinetic energies can be standardized in nuclear reaction databases as \cite{Herman2009}
\begin{eqnarray}\label{eq7}
\frac{d\sigma}{d\Omega^c_{X}}=\sum_l\frac{2l+1}{4\pi}f_l^c(X)P_l(\cos\theta^c_{X}).
\end{eqnarray}
Here, $X=m_1$ or $M_1$. The Legendre expansion coefficient $f^c_l(m_1)$ and $f^c_l(M_1)(=(-1)^lf^c_l(m_1))$ can be derived from the the generilized master equation of the exciton model \cite{Zhang1989,Zhangjs2002}.

Using the non-relativistic triangle relationship of the velocity vectors, the average kinetic energy of $m_1$ in the laboratory system (LS, denoting superscript $l$) can be obtained
\begin{eqnarray}\label{eq8}
\overline{\varepsilon}^l_{m_1} &=& \int \frac{1}{2}m_1(\textbf{\textit{V}}_C+\textbf{\textit{v}}^c_{m_1})^2\frac{d\sigma}{d\Omega^c_{m_1}}d\Omega^c_{m_1}  \nonumber\\
 &=& \frac{m_1m_0E_L}{M^2_C}+\varepsilon^c_{m_1}+\frac{2}{M_C}\sqrt{m_0m_1E_L\varepsilon^c_{m_1}}f^c_1(m_1),
\end{eqnarray}
where $\textbf{\textit{V}}_C$ and $\textbf{\textit{v}}^c_{m_1}$ are the velocity vectors of the mass center and $m_1$ in CMS, respectively. $E_L$ and $m_0$ are the kinetic energy and mass of the incident particle, respectively. Similarly as Eq. (\ref{eq8}), the average kinetic energy of $M_1$ in LS reads
\begin{eqnarray}\label{eq9}
\overline{E}^l_{M_1} = \frac{M_1m_0E_L}{M^2_C}+E^c_{M_1}-\frac{2M_1}{M_C}\sqrt{\frac{m_0E_LE^c_{M_1}}{M_1}}f^c_1(m_1).
\end{eqnarray}

Thus, it is obvious that the energy conservation in the first particle emission process in LS can be strictly kept as follows
\begin{eqnarray}\label{eq10}
\overline{\varepsilon}^l_{m_1}+ \overline{E}^l_{M_1}+E_{k_1}=E_L+B_0-B_1.
\end{eqnarray}
Where, $B_0$ is the binding energy of $m_0$ in $M_C$.

After the first particle $m_1$ is emitted, its residual nucleus $M_1$ at energy level $E_{k_1}$ and with recoiling kinetic energy $E^c_{M_1}$ in CMS will emit the secondary particle $m_2$ with kinetic energy $\varepsilon^c_{m_2}$, if the conservations of the energy, angular momentum and parity are met. Thus the corresponding residual nucleus $M_2$ at energy level $E_{k_2}$ will also gain the recoiling kinetic energy $E^c_{M_2}$ at arbitrary directions in CMS. In order to analytically describe the kinematics of the secondary emitted particle, it is assumed that $M_1$ is static in the recoil nucleus system (RNS, denoting superscript $r$), then the definitive kinetic energy of $m_2$ can be expressed as
\begin{eqnarray}\label{eq11}
\varepsilon^r_{m_2}=\frac{M_2}{M_1}(E_{k_1}-B_2-E_{k_2}).
\end{eqnarray}

Similarly, the energy of $M_2$ in RNS can be also obtained
\begin{eqnarray}\label{eq12}
E^r_{M_2}=\frac{m_2}{M_1}(E_{k_1}-B_2-E_{k_2}).
\end{eqnarray}

\begin{figure}
\centering
\includegraphics[width=8cm,angle=0]{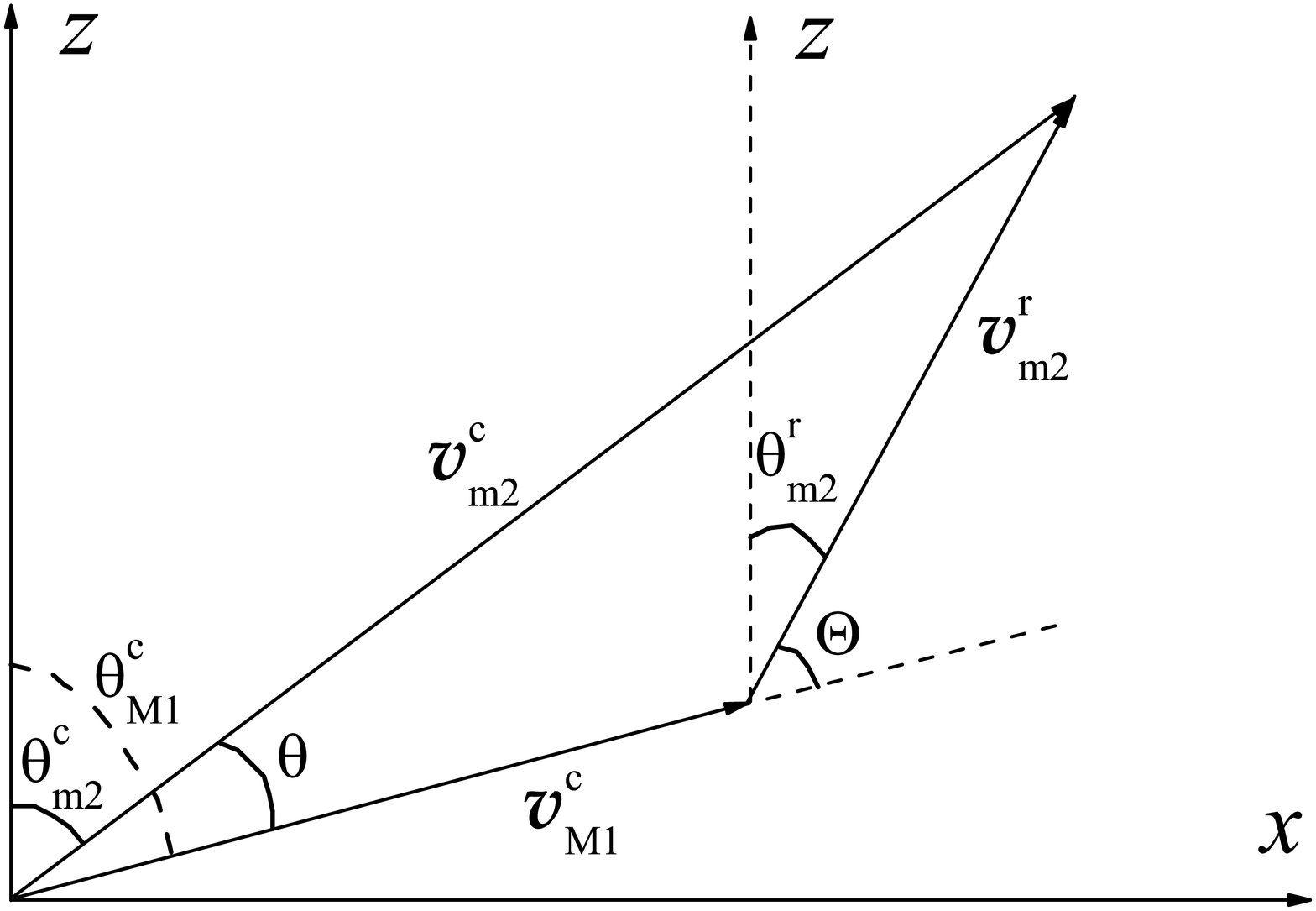}
\caption{Triangle relationship of the velocity vectors $\textbf{\textit{v}}^c_{m_2}, \textbf{\textit{v}}^c_{M_1}$ and $\textbf{\textit{v}}^r_{m_2}$ in $x-z$ plane.}\label{fig1}
\end{figure}

Using the non-relativistic triangle relationship $\textbf{\textit{v}}^c_{m_2}=\textbf{\textit{v}}^c_{M_1}+\textbf{\textit{v}}^r_{m_2}$ as shown in Fig. \ref{fig1}, we can get \cite{Zhang1999C12,Zhang2003B11}
\begin{eqnarray}\label{eq13}
\varepsilon^c_{m_2}=\varepsilon^r_{m_2}(1+2\gamma\cos\Theta+\gamma^2),
\end{eqnarray}
\begin{eqnarray}\label{eq14}
\cos\Theta=\sqrt{\frac{\varepsilon^c_{m_2}}{\varepsilon^r_{m_2}}}[\cos\theta^c_{m_2}\cos\theta^c_{M_1}+\sin\theta^c_{m_2}\sin\theta^c_{M_1}\cos(\varphi^c_{m_2}-\varphi^c_{M_1})]-\gamma,
\end{eqnarray}
where $\gamma \equiv \sqrt{\frac{m_2E^c_{M_1}}{M_1\varepsilon^r_{m_2}}}$. The maximum and minimum kinetic energies of $m_2$ in CMS are given by the following
\begin{eqnarray}\label{eq15}
\varepsilon^c_{m_2,max}=\varepsilon^r_{m_2}(1+\gamma)^2, ~~~~\varepsilon^c_{m_2,min}=\varepsilon^r_{m_2}(1-\gamma)^2.
\end{eqnarray}

In the frame of STLN, the double-differential cross section of $m_2$ in RNS is assumed as the following isotropic distribution with a definitive kinetic energy $\varepsilon^r_{m_2}$, i.e.,
\begin{eqnarray}\label{eq16}
\frac{d^2\sigma}{d\varepsilon^r_{m_2}d\Omega^r_{m_2}}=\frac{1}{4\pi}\delta[\varepsilon^c_{m_2}-\varepsilon^r_{m_2}(1+2\gamma\cos\Theta+\gamma^2)].
\end{eqnarray}

Starting from the basic relation of the double-differential cross sections between CMS and RNS, the double-differential cross section of $m_2$ in CMS can be obtained through the corresponding results in RNS averaged by the angular distribution of $M_1$, i.e.,
\begin{eqnarray}\label{eq17}
\frac{d^2\sigma}{d\varepsilon^c_{m_2}d\Omega^c_{m_2}}=\int d\Omega^c_{M_1} \frac{d\sigma}{d\Omega^c_{M_1}} \sqrt{\frac{\varepsilon^c_{m_2}}{\varepsilon^r_{m_2}}} \frac{d^2\sigma}{d\varepsilon^r_{m_2}d\Omega^r_{m_2}}.
\end{eqnarray}

By means of the properties of $\delta$ function and Eqs. (\ref{eq7})-(\ref{eq17}),  the double-differential cross section of $m_2$ in CMS can be rewritten as \cite{Zhang2015}
\begin{eqnarray}\label{eq18}
\frac{d^2\sigma}{d\varepsilon^c_{m_2}d\Omega^c_{m_2}}=\frac{1}{16\pi^2\gamma\varepsilon^r_{m_2}}\sum_l (2l+1)f_l^c(M_1)\int_0^{\pi} dtP_l(\sqrt{(1-\eta^2)\sin^2\theta^c_{m_2}}\cos t+\eta\cos\theta^c_{m_2}),         \nonumber\\
\end{eqnarray}
where $\eta=\sqrt{\frac{\varepsilon^r_{m_2}}{\varepsilon^c_{m_2}}}\frac{\varepsilon^c_{m_2}/\varepsilon^r_{m_2}-1+\gamma^2}{2\gamma}$.
With the new integral formula introduced in Set. \ref{sect2}, Eq. (\ref{eq18}) can be simplified as follows
\begin{eqnarray}\label{eq19}
\frac{d^2\sigma}{d\varepsilon^c_{m_2}d\Omega^c_{m_2}}=\sum_l \frac{(-1)^l}{16\pi\gamma\varepsilon^r_{m_2}}(2l+1)f_l^c(m_1)P_l(\eta)P_l(\cos\theta^c_{m_2}).
\end{eqnarray}

It is worth mentioning that Eqs. (\ref{eq18}) and (\ref{eq19}) are realized only for nuclear reactions without polarization of incoming nucleons and orientation of target nuclei, which are also the constraint conditions of STLN. Therefore, the integral formula Eq. (\ref{eq4}) can not be applied to the reactions with polarization of incoming nucleons and orientation of target nuclei.

The normalized double-differential cross section of the secondary emitted particle $m_2$ is also standardized in nuclear reaction databases as \cite{Herman2009}
\begin{eqnarray}\label{eq20}
\frac{d^2\sigma}{d\varepsilon^c_{m_2}d\Omega^c_{m_2}}=\sum_l\frac{2l+1}{4\pi}f_l^c(m_2)P_l(\cos\theta^c_{m_2}).
\end{eqnarray}

By comparing Eqs. (\ref{eq19}) and (\ref{eq20}), the Legendre expansion coefficients of $m_2$ in CMS can be expressed as
 \begin{eqnarray}\label{eq21}
f_l^c(m_2)=\frac{(-1)^l}{4\gamma\varepsilon^r_{m_2}}f_l^c(m_1)P_l(\eta).
\end{eqnarray}

Similarly as Eq. (\ref{eq21}), we can also derive the analytical Legendre expansion coefficients of $M_2$ in CMS as
 \begin{eqnarray}\label{eq22}
f_l^c(M_2)=\frac{(-1)^l}{4\Gamma E^r_{M_2}}f_l^c(m_1)P_l(H),
\end{eqnarray}
where $\Gamma=\sqrt{\frac{M_2 E^c_{M_1}}{M_1E^r_{M_2}}}$ and $H=\sqrt{\frac{E^r_{M_2}}{E^c_{M_2}}}\frac{E^c_{M_2}/E^r_{M_2}-1+\Gamma^2}{2\Gamma}$.

It is obvious that the Legendre expansion coefficients of $m_2$ and $M_2$ in CMS are closely related to $m_1$ and $M_1$. Analytical expressions of Eq. (\ref{eq21}) and (\ref{eq22}) can largely reduce the volume of file-6 in nuclear reaction databases.

In CMS, the average kinetic energy of $m_2$ can be obtained by averaging its double differential cross section, i.e.,
\begin{eqnarray}\label{eq23}
\overline{\varepsilon}^c_{m_2} &=& \int^{\varepsilon^c_{m_2,max}}_{\varepsilon^c_{m_2,min}}\varepsilon^c_{m_2}
\frac{d^2\sigma}{d\varepsilon^c_{m_2}d\Omega^c_{m_2}}d\varepsilon^c_{m_2}d\Omega^c_{m_2}  \nonumber\\
&=& \varepsilon^r_{m_2}(1+\gamma^2).
\end{eqnarray}

We also can obtain the average kinetic energy of $M_2$ in CMS in the same way, i.e.,
\begin{eqnarray}\label{eq24}
\overline{E}^c_{M_2}= E^r_{M_2}(1+\Gamma^2).
\end{eqnarray}

In terms of the non-relativistic triangle relationship of the velocity vectors, the average kinetic energy of $m_2$ in LS can be obtained
\begin{eqnarray}\label{eq25}
\overline{\varepsilon}^l_{m_2} &=& \int \frac{1}{2}m_2(\textbf{\textit{V}}_C+\textbf{\textit{v}}^c_{m_2})^2\frac{d^2\sigma} {d\varepsilon^c_{m_2}d\Omega^c_{m_2}}d\varepsilon^c_{m_2}d\Omega^c_{m_2}  \nonumber\\
 &=& \frac{m_0m_2E_L}{M^2_C}+\overline{\varepsilon}^c_{m_2}-2\frac{m_2}{M_C} \sqrt{\frac{m_0E_L E^c_{M_1}}{M_1}}f^c_1(m_1).
\end{eqnarray}

In the same way, the average kinetic energy of $M_2$ in LS can be derived as
\begin{eqnarray}\label{eq26}
\overline{E}^l_{M_2} = \frac{m_0M_2E_L}{M^2_C}+\overline{E}^c_{M_2}-2\frac{M_2}{M_C} \sqrt{\frac{m_0E_L E^c_{M_1}}{M_1}}f^c_1(m_1).
\end{eqnarray}

Thus, the energy conservation of the initial and final states in the sequential secondary particle emission processes for the light nucleus reactions can be strictly kept in LS as follows
\begin{eqnarray}\label{eq27}
E^l_{total} &=& \overline{\varepsilon}^l_{m_1}+\overline{\varepsilon}^l_{m_2}+ \overline{E}^l_{M_2}+E_{k_2} \nonumber\\
 &=& E_L+B_0-B_1-B_2.
\end{eqnarray}

After the first particle emission, the residual nucleus $M_1$ at energy level $E_{k_1}$ is possible to spontaneously break up into two smaller particles or nuclei. For example, in neutron induced $^7$Li reaction \cite{Zhang2002Li7}, the compound nucleus $^8$Li with excited energy $E^*$ will emit a triton, and the corresponding residual nucleus $^5$He is unstable which can spontaneously break up into a neutron and an alpha. Specially, there are double two-body breakup processes in light nucleus reactions. In the case of neutron induced $^9$Be reaction \cite{Sun2009Be9,Duan2010Be9}, the compound nucleus $^{10}$Be with excited energy $E^*$ can emit two $^5$He, and each $^5$He can spontaneously break up into a neutron and an alpha. The double-differential cross sections of these reaction processes can be derived by using the same approach as introduced above. In addition, there are some reaction processes in which the residual nucleus $M_2$ can either emit the third particle or break up into smaller particles spontaneously. For example, reaction channels $^{12}$C(n, n$\alpha$)$^8$Be and $^{12}$C(n, 2$\alpha$)$^5$He, are important channels in neutron induced $^{12}$C reaction \cite{Sun2007C12,Sun2008kerma}. These reaction processes are the universal phenomena in neutron or proton induced 1p-shell light nucleus reactions, which are effected by the emissions of the first and secondary particles and can be well described by STLN. In other words, the integral formula Eq. (\ref{eq4}) and the Legendre expansion coefficients Eq. (\ref{eq21}) of the secondary emitted particle $m_2$ can be further used to describe these sequential emission processes and multi-particle productions from spontaneous breakup processes.

\begin{figure}
\centering
\includegraphics[width=10cm,angle=0]{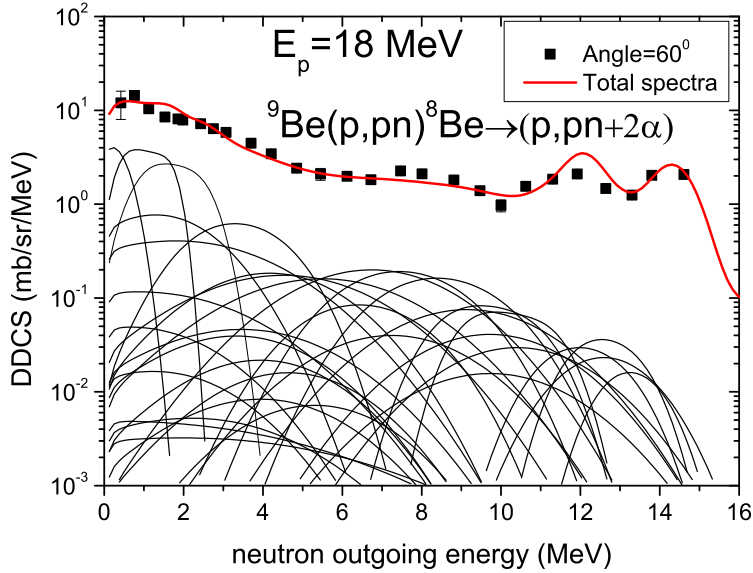}
\caption{(Color online) The partial energy-angular spectra of (p, pn)$^8$Be$\rightarrow$(p, pn+2$\alpha$) reaction with outgoing angle 60$^0$ at $E_p$=18 MeV in LS. The points denote the experimental data taken from Ref. \cite{Verbinski1969}, and the red solid line denotes the calculated total outgoing neutron energy-angular spectra. The black solid lines denote the partial neutron spectra coming from the 1st-17th excited energy levels of the first residual nucleus $^9$Be to the lowest five energy levels of the secondary residual nucleus $^8$Be. Only the cross sections with the values larger than 0.1 mb are given.}\label{fig2}
\end{figure}

\subsection{Applications to nucleon induced light nucleus reactions}
For neutron induced 1p-shell light nuclei, such as $^{6}$Li \cite{Zhang2001Li6}, $^{7}$Li \cite{Zhang2002Li7}, $^9$Be \cite{Sun2009Be9,Duan2010Be9}, $^{10}$B \cite{Zhang2003B10}, $^{11}$B \cite{Zhang2003B11}, $^{12}$C \cite{Zhang1999C12,Sun2008kerma}, $^{14}$N \cite{Yan2005N14} and $^{16}$O \cite{Zhang2001O16,Duan2005O16}, the calculated double-differential neutron-production cross sections agree greatly well with the experimental data. In these reactions, a large part of the contributions to the total energy-angular spectra comes from the secondary emitted neutrons. Therefore, the integral formula (\ref{eq4}) and the Legendre expansion coefficients Eq. (\ref{eq21}) of the secondary emitted particle are very important for the calculations of the partial energy-angular spectra of the sequential emission processes.

\begin{figure}
\centering
\includegraphics[width=10cm,angle=0]{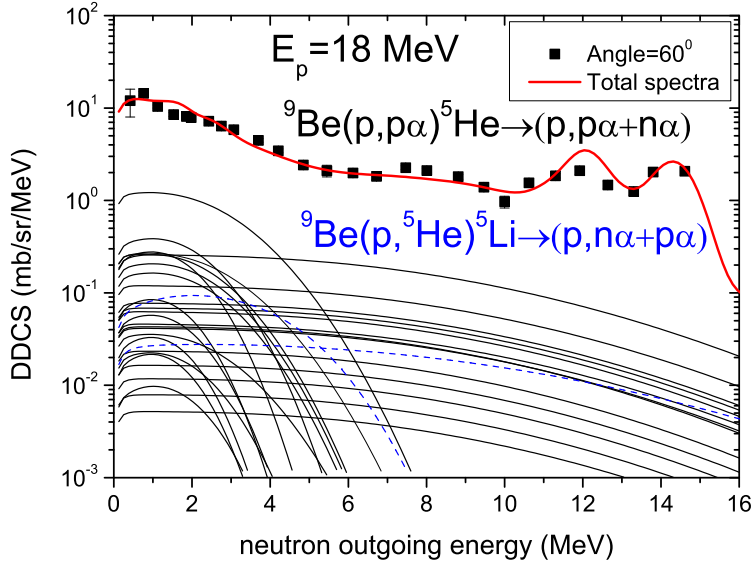}
\caption{(Color online) The same as Fig. \ref{fig2}. But the black solid lines denote the partial spectra of the emitted neutron from (p, p$\alpha$)$^5$He$\rightarrow$(p, p$\alpha$+n$\alpha$) reaction. The contributions of these partial neutron spectra come from the 4th-17th excited energy levels of the first residual nucleus $^9$Be to the lowest two energy levels of the secondary residual nucleus $^5$He, which can spontaneously breakup a neutron and an alpha. And the blue dash lines denote the partial spectra of the emitted neutron from (p, $^5$He)$^5$Li$\rightarrow$(p, n$\alpha$+p$\alpha$) reaction. The contributions of these partial neutron spectra come from the ground state and the 1st excited energy level of $^5$He, too.}\label{fig3}
\end{figure}

 In this section, we will extend the integral formula (\ref{eq4}) and the Legendre expansion coefficients Eq. (\ref{eq21}) to describe the secondary emitted particles in proton induced light nucleus reactions \cite{Zhang2015}. In the case of p+$^9$Be reaction at $E_p$=18 MeV with outgoing angle 60$^0$, the double-differential neutron-production cross sections of the reaction channel (p, pn)$^8$Be$\rightarrow$(p, pn+2$\alpha$) are shown by these black solid lines in Fig. \ref{fig2}. These black solid lines denote the contributions of the partial neutron spectra from the 1st-17th excited energy levels of the first residual nucleus $^9$Be to the lowest five energy levels of the secondary residual nucleus $^8$Be, where the conservations of the energy, angular momentum and parity are strictly kept. In Fig. \ref{fig3}, the black solid lines denote the partial neutron spectra from the reaction channel (p, p$\alpha$)$^5$He$\rightarrow$(p, p$\alpha$+n$\alpha$). The contributions of these partial neutron spectra come from the 4th-17th excited energy levels of the first residual nucleus $^9$Be to the lowest two energy levels of the secondary residual nucleus $^5$He, which can spontaneously break up into a neutron and an alpha. And the blue dash lines denote the partial neutron spectra from the reaction channel (p, $^5$He)$^5$Li$\rightarrow$(p, n$\alpha$+p$\alpha$). However, the contributions of these partial neutron spectra come from the ground state and the 1st excited energy level of $^5$He, which is regarded as the first residual nucleus.

\begin{figure}
\centering
\includegraphics[width=10cm,angle=0]{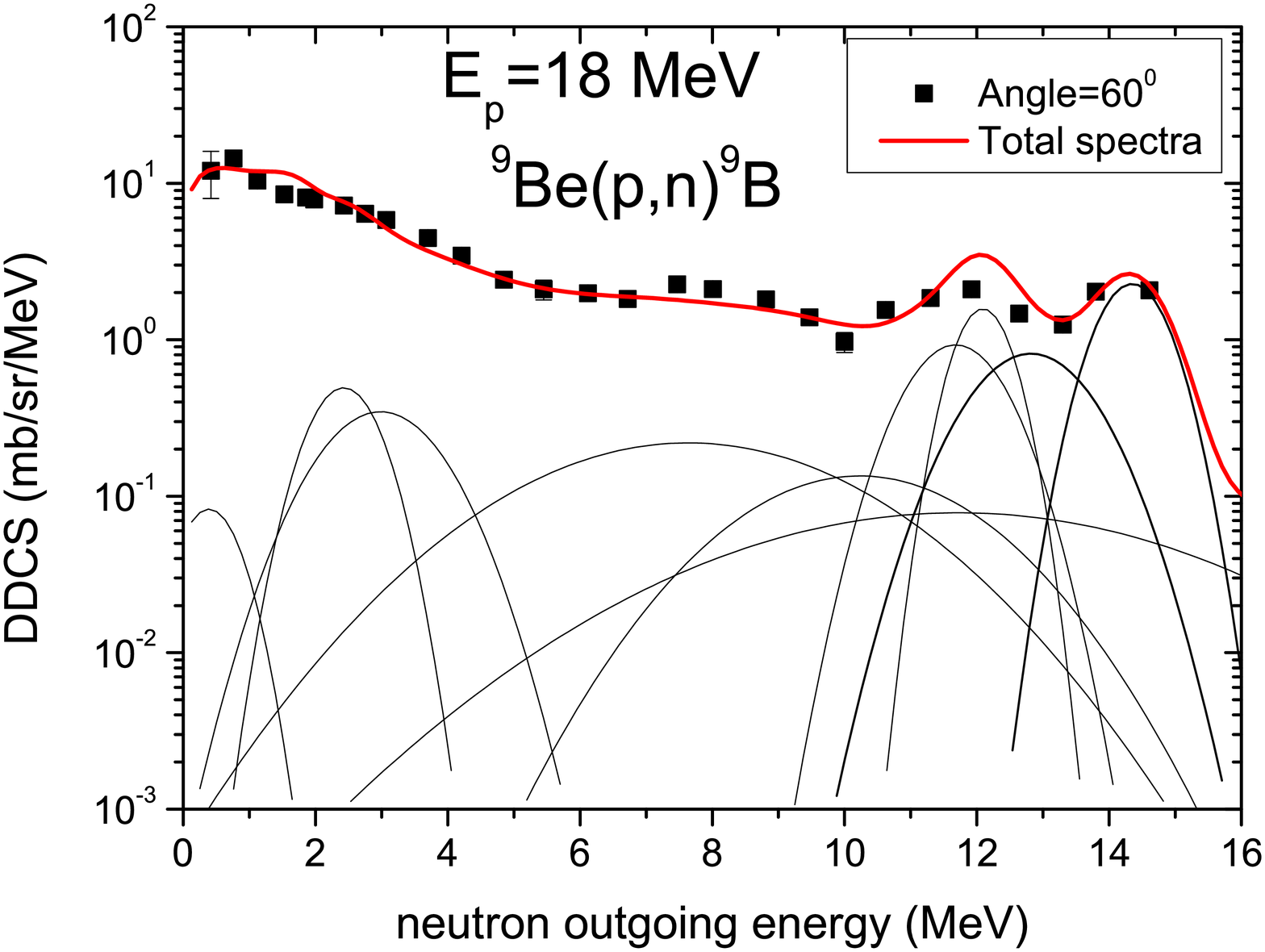}
\caption{(Color online) The same as Fig. \ref{fig2}. But the black solid lines denote the partial spectra of the first emitted neutron from (p, n)$^9$B reaction. The contributions of these partial neutron spectra, in which broadening effects must be taken into account, come from the ground state to 9th excited energy levels of $^9$B.}\label{fig4}
\end{figure}

 Besides three sequential multi-particle emission reaction channels as shown in Figs. \ref{fig2} and \ref{fig3}, another important reaction channel is the first neutron emission, i.e., (p, n)$^9$B reaction channel as shown in Fig. \ref{fig4}. The contributions of the first emitted neutron, of which broadening effects must be taken into account \cite{Zhang1999C12}, come from the ground state to 9th excited energy levels of $^9$B. Summing up all of outgoing neutron partial energy-angular spectra (black solid and blue dash lines in Figs. \ref{fig2}-\ref{fig4}), we can obtain the total outgoing neutron energy-angular spectra as shown by the red line. In these figures, the points denote the experimental data measured by Verbinski et al in 1969 \cite{Verbinski1969}. One can see that the calculated total energy-angular spectra agree greatly well with the experimental data. Furthermore, in the frame of STLN, the calculated total energy-angular spectra at other outgoing angles also agree well with the experimental data \cite{Zhang2015}.

\section{Summary}
\label{sect4}
Since the light nucleus reactions have very strong recoil motion, the particles emitted from the residual nucleus have a very strong backward tendency, while the first emitted particles have an obvious forward tendency. For analytically describing this recoil effects and the energy-angular spectra of the sequentially emitted particles, a new integral formula, which has not been compiled in any integral tables or mathematical softwares, has been employed to obtain analytical Legendre expansion coefficients of the secondary emitted particles. The partial energy-angular spectra of the secondary emitted particles between the discrete energy levels of the residual nuclei are very important parts of the total energy-angular spectra. In the calculations of neutron induced 1p-shell light nucleus reactions, the total double-differential neutron cross sections agree fairly well with the measured data. Taking p+$^9$Be reaction at $E_p$=18 MeV as an example, one can see that the total double-differential neutron cross sections also agree well with the measured data. We expect that this new integral formula will be extend to all of the 1p-shell light nucleus reactions induced by nucleon (including neutron and proton) in the frame of STLN.
So this integral formula can largely reduce the volume of file-6 in nucleon induced nuclear reaction databases with full energy balance. Therefore, this integral formula and STLN are being tested by proton induced other light elements.

\textbf{Acknowledgements}
We wish to thank Dr. Guo Yun, professors Wang Ning, Ou Li, Liu Min and anonymous referees for some valuable suggestions. This work is supported by the National Natural Science Foundation of China (No. 11465005); the Natural Science Foundation of Guangxi (No. 2014GXNSFDA118003); Guangxi University Science and Technology Research Project (No. 2013ZD007); the Open Project Program of State Key Laboratory of Theoretical Physics, Institute of Theoretical Physics, Chinese Academy of Sciences, China (No. Y4KF041CJ1);
and the project of outstanding young teachers' training in higher education institutions of Guangxi.


\begin{references}

\bibitem{Herman2009}Trkov A, Herman M, and Brown D A. \textit{ENDF-6 Formats Manual}. Brookhaven National Laboratory, Upton, NY, USA, 11973-5000 (2011).

\bibitem{FENDL-3.02015}IAEA. \textit{Fusion Evaluated Nuclear Data Library}, 2015. See  \href{https://www-nds.iaea.org/fendl30/}{https://www-nds.iaea.org/fendl30/}.

\bibitem{TENDL-2014}Koning A J, Rochman D, van der Marck SC, et al. \textit{TALYS-based Evaluated Nuclear Data Library}, 2014. See \href{ftp://ftp.nrg.eu/pub/www/talys/tendl2014/tendl2014.html/}{ftp://ftp.nrg.eu/pub/www/talys/tendl2014/tendl2014.html/}.

\bibitem{JEFF-3.22014}OECD Nuclear Energy Agency. \textit{The Joint Evaluated Fission and Fusion File}, 2014. See \href{http://www.oecd-nea.org/dbforms/data/eva/evatapes/jeff-32/}{http://www.oecd-nea.org/dbforms/data/eva/evatapes/jeff-32/}.

\bibitem{JENDL-4.0u22012}Japan Atomic Energy Agency. \textit{Japanese evaluated nuclear data library}, 2012. See \href{http://wwwndc.jaea.go.jp/jendl/j40/update/}{http://wwwndc.jaea.go.jp/jendl/j40/update/}.

\bibitem{ENDF/B-VII.12011}National Nuclear Data Centre. \textit{US Evaluated Nuclear Data Library}, 2011. See \href{http://www.nndc.bnl.gov/exfor/endfb7.1.jsp}{http://www.nndc.bnl.gov/exfor/endfb7.1.jsp}.

\bibitem{ROSFOND-2010}Russia. \textit{Issued neutron library}, 2010. See \href{http://www.ippe.ru/podr/abbn/libr/rosfond.php}{http://www.ippe.ru/podr/abbn/libr/rosfond.php}.

\bibitem{Zhang2011}Zhang J S, Han Y L and Duan J F, \textit{Journal of the Korean Physical Society} \textbf{59}, 843 (2011).

\bibitem{Zhang2009} Zhang J S, \textit{Statistical Theory of Neutron Induced Reactions of Light Nuclei (Second Edition, in Chinese)}. Science Press, Beijing, (2015).

\bibitem{Zhang2015} Zhang J S, Sun X J, \textit{Statistical Theory of Light Nucleus Reactions (in Chinese)}. private communication, (2015).

\bibitem{Gradshteyn2007}Gradshteyn I S, Ryzhik I M, \textit{Table of Integrals, Series, and Products (Seventh Edition)}. University of Newcastle upon Tyne, England, (2007).

\bibitem{Arad2002}Arad Z, Muzychuk M, \textit{Standard integral table algebras generated by a non-real element of small degree}. Springer, (2002).

\bibitem{Bradley1995}Bradley G L, Smith K J, \textit{Student mathematics handbook and integral table for calculus}. Prentice-Hall, Inc, (1995).

\bibitem{Mathematica}Wolfram Research, \href{http://www.wolfram.com/}{http://www.wolfram.com/}.

\bibitem{Matlab}MathWorks, \href{http://www.mathworks.com/products/matlab/index-b.html}{http://www.mathworks.com/products/matlab/index-b.html}.

\bibitem{Maple}Maplesoft, \href{http://www.maplesoft.com/}{http://www.maplesoft.com/}.

\bibitem{Zhang1989}Zhang J S, Shi X, The formulation of UNIFY code for the calculation of fast neutron data for structural materials, INDC(CPR)-014, (1989).

\bibitem{Zhangjs2002}Zhang J S, \textit{Nucl. Sci. Eng.} \textbf{142}, 207 (2002).

\bibitem{Zhang1999C12}Zhang J S, et al, \textit{Nucl. Sci. Eng.} \textbf{133}, 218 (1999).

\bibitem{Zhang2003B11}Zhang J S, \textit{Commun. Theor. Phys.} \textbf{39}, 83 (2003).

\bibitem{Zhang2002Li7}Zhang J S, HAN Y L, \textit{Commun. Theor. Phys.} \textbf{37}, 465 (2002).

\bibitem{Sun2009Be9}Duan J F, Zhang J S, Wu H C and Sun X J, \textit{Phys. Rev. C} \textbf{80}, 064612 (2009).

\bibitem{Duan2010Be9}Duan J F, Zhang J S, Wu H C and Sun X J, \textit{Commun. Theor. Phys.} \textbf{54}, 129 (2010).

\bibitem{Sun2007C12}Sun X J, Duan J F, Wang J M and Zhang J S, \textit{Commun. Theor. Phys.} \textbf{48}, 534 (2007).

\bibitem{Sun2008kerma}Sun X J, Qu W J. Duan J F and Zhang J S, \textit{Phys. Rev. C} \textbf{78}, 054610 (2008).

\bibitem{Zhang2001Li6}Zhang J S, \textit{Commun. Theor. Phys.} \textbf{36}, 437 (2001).

\bibitem{Zhang2003B10}Zhang J S, \textit{Commun. Theor. Phys.} \textbf{39}, 433 (2003).

\bibitem{Yan2005N14}Yan Y L, Duan J F, Sun X J, et al, \textit{Commun. Theor. Phys.}, \textbf{44}, 128 (2005).

\bibitem{Zhang2001O16}Zhang J S, et al, \textit{Commun. Theor. Phys.} \textbf{35}, 579 (2001).

\bibitem{Duan2005O16}Duan J F, Yan Y L, et al, \textit{Commun. Theor. Phys.} \textbf{44}, 701 (2005).

\bibitem{Verbinski1969}Victor V. Verbinski, Walter R. Burrus, \textit{Phys. Rev. C} \textbf{177}, 1671 (1969).

\end{references}
\end{document}